\documentclass[twocolumn,showpacs,preprintnumbers,amsmath,amssymb]{revtex4}
\usepackage{graphicx}
\usepackage{dcolumn}
\usepackage{bm}
\usepackage{natbib}

\def\ltsima{$\; \buildrel < \over \sim \;$}
\def\simlt{\lower.5ex\hbox{\ltsima}}
\def\gtsima{$\; \buildrel > \over \sim \;$}
\def\simgt{\lower.5ex\hbox{\gtsima}}

\begin{document}

\title{Observing Baryon Oscillations with Cosmic Shear}

\author{Fergus Simpson}
\email{frgs@ast.cam.ac.uk} \affiliation{Institute of Astronomy,
University of Cambridge, Madingley Road, Cambridge CB3 0HA}

\date{\today}
\newcommand{\ud}{\mathrm{d}}

\begin{abstract}

A cosmic shear survey, spanning a significant proportion of the
sky, should greatly improve constraints on a number of
cosmological parameters. It also provides a unique opportunity to
examine the matter power spectrum directly. However, the observed
lensing signal corresponds to a weighted average of the power
spectrum across a range of scales, and so the potential to resolve
the baryon oscillations has been somewhat neglected. These
features originated prior to recombination, induced by the
acoustics of the photon-baryon fluid.  Recent galaxy surveys have
detected the imprints
\cite{2005astro.ph..1171E,2005astro.ph..1174C}, and in the future
such measurements may even be used to refine our understanding of
dark energy.

Without redshift information, cosmic shear is an ineffective probe
of the baryon oscillations. However, by implementing a novel
\emph{multipole-dependent} selection of photometric redshift bins,
sensitivity is improved by an order of magnitude, bringing the
``wiggles" within reach of future surveys. As an illustration, we
show that data from surveys scheduled within the next ten years
will be able to distinguish a smoothed power spectrum at the
2$\sigma$ level.

\end{abstract}
 \maketitle

\section{Introduction}
Consider the observation of a pair of distant galaxies. Their
images accumulate a correlated distortion as the light is
gravitationally deflected throughout the multi-billion year
journey, thereby generating an illusion of alignment. The extent
of this lensing signal, known as cosmic shear, is related to the
magnitude of density perturbations lying between the two paths.
Thus as their routes gradually converge, the sensitivity to the
matter power spectrum traces an ever-decreasing distance scale.
Since the power spectrum is probed over such a broad range, it was
not anticipated that cosmic shear surveys could resolve any
oscillatory features.

However, the cosmic shear technique offers much promise, with
ambitious surveys planned which far surpass those that have been
conducted to date. These will push various cosmological parameters
to sub-percent levels of precision, provided systematic effects
can be sufficiently controlled. It also presents the most direct
probe of the matter power spectrum, and recently particular
emphasis has been placed on constraining the dark energy equation
of state $w$
\cite{2004PhRvD..70f3510S,2005PhRvD..71h3501S,2004AJ....127.3102R,2002PhRvD..65b3003H}.

In this work we aim to show how the presence of baryon
oscillations can be inferred from cosmic shear data, acting as an
auxiliary probe to galaxy redshift surveys. The importance of
independent evidence cannot be underestimated, as it is essential
if we are to claim a concordant model. A successful detection will
also build confidence in the validity of other parameters.

Central to our approach is the concept of dividing the galaxies
into different redshift bins when analysing different angular
separations. This allows us to compare the shear signal which is
suppressed by the wiggles, with those which are enhanced.

In \S II we briefly review the theory of cosmic shear, and outline
the fiducial survey. \S III then determines the redshift bins we
will use to reveal the baryon oscillations, using the optimised
statistic defined in \S IV. The errors associated with this
statistic are derived in \S V.

\section{The Fiducial Cosmic Shear Survey}

For our purposes, there are two important inputs which determine
the form of the cosmic shear signal - the matter power spectrum,
and the distribution of source galaxies. Essentially, we will look
to optimally probe the former, by manipulating the latter. The
cosmic shear power spectrum is given by

\begin{equation} \label{eq:shear}
C_\ell = \frac{\displaystyle 9}{16}\Big(\frac{\displaystyle
H_0}{\displaystyle c}\Big)^4 \Omega_m^2 \int
\Big[\frac{\displaystyle g(\chi)}{\displaystyle
ar(\chi)}\Big]^2P(\frac{\ell}{r},\chi) \ud \chi ,
\end{equation}

\noindent where $\chi$ is the coordinate distance, and $r(\chi)$
the comoving angular diameter distance. The galaxy distribution
$n(\chi)$ features within the lensing efficiency $g$, given by

\begin{equation}\label{eq:lenseff}
g(\chi)=2 \int_{\chi}^{\chi_h}
n(\chi')\frac{r(\chi)r(\chi'-\chi)}{r(\chi')} \ud\chi' ,
\end{equation}
where
\begin{equation}\label{eq:normal}
n(z) \propto z^\alpha e^{-(z/z_0)^\beta} .
\end{equation}
The constants $\alpha$, $\beta$, and $z_0$ are taken from
Refregier et al. \cite{2004AJ....127.3102R}. We will assume a flat
universe throughout, and hence $r(\chi)=\chi$.

The survey parameters outlined in Table~\ref{tab:survey} are
adopted in order to emulate the performance of next-generation
surveys currently in the early development phase, such as
LSST\footnote{http://www.lsst.org/},
SNAP\footnote{http://snap.lbl.gov/} and DUNE.

Now consider the error of a given multipole. The standard formula
for the error of an autocorrelation is given by Kaiser
\cite{1998ApJ...498...26K},

\begin{equation}
\label{eq:error}
\sigma(\ell)= \sqrt{\frac{\displaystyle 2}{\displaystyle(2\ell
+1)f_{sky}}}(C_\ell+\frac{\displaystyle\sigma^2_\gamma}{\displaystyle
2n_g}) ,
\end{equation}

\noindent where $n_g$ denotes the number of galaxies per steradian
within the bin, and $f_{sky}$ is the fraction of sky covered by the
survey. The rms prelensing ellipticity of sources is given by
$\sigma_{\gamma}$.

Cosmological parameters are summarised in Table~\ref{tab:cosmo}.
CAMB\footnote{http://camb.info/} is used to produce the matter
power spectrum $P(k,z)$.

\begin{table} [t]
\caption{\label{tab:survey}Parameters adopted for the fiducial
survey.}
\begin{tabular}[b]{c|c|c|c|c|c} 
 \hline
 $\alpha$ & $\beta$ & $z_0$ & Area    & $\bar{n}$ & $\sigma_\gamma$\\
          &       &         & ($deg^2$)&  $(arcmin^{-2})$ &          \\
  \hline
 2        &     2   &   1.13  &  20,000 &    100  &  0.2\\
\end{tabular}
\end{table}

\begin{table} [b]
\caption{\label{tab:cosmo}The fiducial $\Lambda$CDM model.}
\begin{tabular}[b]{c|c|c|c|c|c}  
\hline
 $\Omega_m$ & $\sigma_8$ & $h$ & w    & $\Omega_\Lambda$ & $\Omega_b$ \\
 \hline
 0.3       &     0.88   &   0.7   &  -1 &    0.7         &   0.0462  \\
\end{tabular}
\end{table}

\begin{figure}
\includegraphics[width=3.4in]{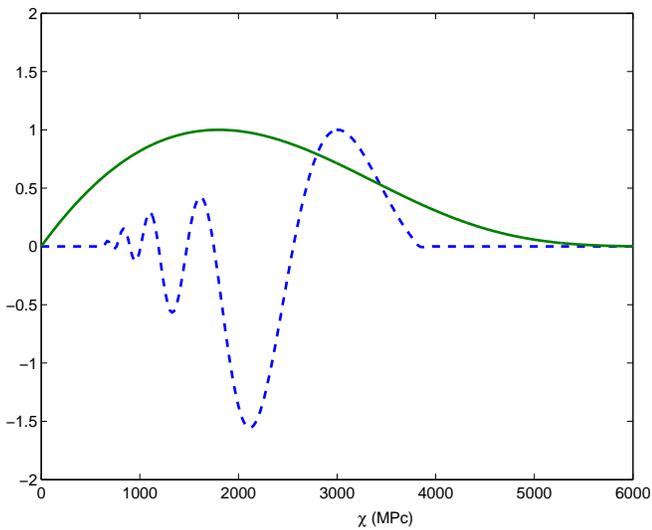}
\caption{\label{fig:smooth} This illustrates the difficulty of
differentiating between a smooth and oscillatory power spectrum.
The strength of the signal is approximately related to the
integral of the product of the two lines. The solid line
represents $g$, the lensing efficiency, whilst the dashed line is
given by $P(k)-P(k)^{smooth}$ at $\ell=150$. }
\end{figure}

The challenge of distinguishing the ``wiggles" becomes apparent
when considering the plot of Fig.~\ref{fig:smooth}. Roughly
speaking, the deviation produced by their presence is related to
the integral of the product of the two lines. Thus if we could
generate a similar oscillation in the (solid line), a significant
improvement could be made. Fortunately, we do have some control
over $g$.

\section{Tomography}

By dividing the source galaxies into redshift bins, we generate
different power spectra. The key to revealing the baryon
oscillations from within the cosmic shear signal lies in comparing
the correlations of galaxies which have their shear enhanced by
the ``wiggles" with those that are suppressed. The focus will be
on angular scales corresponding to $100 \lesssim \ell \lesssim
300$, since these multipoles are lensed by the first few
oscillations in the power spectrum.

As outlined in Eisenstein \& Hu \cite{1998ApJ...496..605E}, the
form of the oscillatory envelope in the power spectrum is well
described by $ j_0(k\tilde{s})$ where the bessel function $j_0(x)$
is defined as $\sin{x}/x$, $\tilde{s}$ is the effective sound
horizon at the drag epoch, and for our purposes $k=\ell/\chi$ from
(\ref{eq:shear}). We therefore aim to generate a lensing
efficiency $g$ which can tune in to this signal, by careful
selection of our redshift bins.

To reveal how the distribution $n(z)$ influences $g$, we
differentiate (\ref{eq:lenseff}) to find

\begin{equation}
\frac{\ud^3g}{\ud \chi^3} = 2\frac{\ud n(\chi)}{\ud
\chi}+\frac{4n(\chi)}{\chi} .
\end{equation}

By inspection, we anticipate a function of the form $n(\chi) =
\sin(\tilde{s}\ell/\chi+\phi)$ could have the desired effect of
producing a lensing efficiency which traces the oscillations in
the power spectrum. Thus we split our population of galaxies into
two intertwined bins. 

\begin{eqnarray}
\label{eq:na} n_A(\chi,\ell) = \left\{ \begin{array}{ll}
 n(\chi) & \epsilon(\chi,\ell)>0\\
0 & \epsilon(\chi,\ell)<0\\
\end{array}\right.\\
\label{eq:nb}
 n_B(\chi,\ell)=\left\{ \begin{array}{ll}
 0 & \epsilon(\chi,\ell)>0\\
 -n(\chi) & \epsilon(\chi,\ell)<0  \\
 \end{array}\right.
\end{eqnarray}

\noindent where $\epsilon(\chi ,\ell) \equiv
\sin(\tilde{s}\ell/\chi+\phi)$. This is the trial function, which
we hope will divide the galaxy population into bins whose lensing
kernels track the peaks and troughs of the baryon oscillations.
The phase parameter $\phi$ is determined simply by exploring
several values and selecting that which optimises the signal
generated by lensing from the wiggles
$P(k)^{wig}=P(k)^{fid}-P(k)^{smooth}$, i.e. the difference between
a standard and smoothed power spectrum.

\begin{figure} [t]
\includegraphics[width=3.4in]{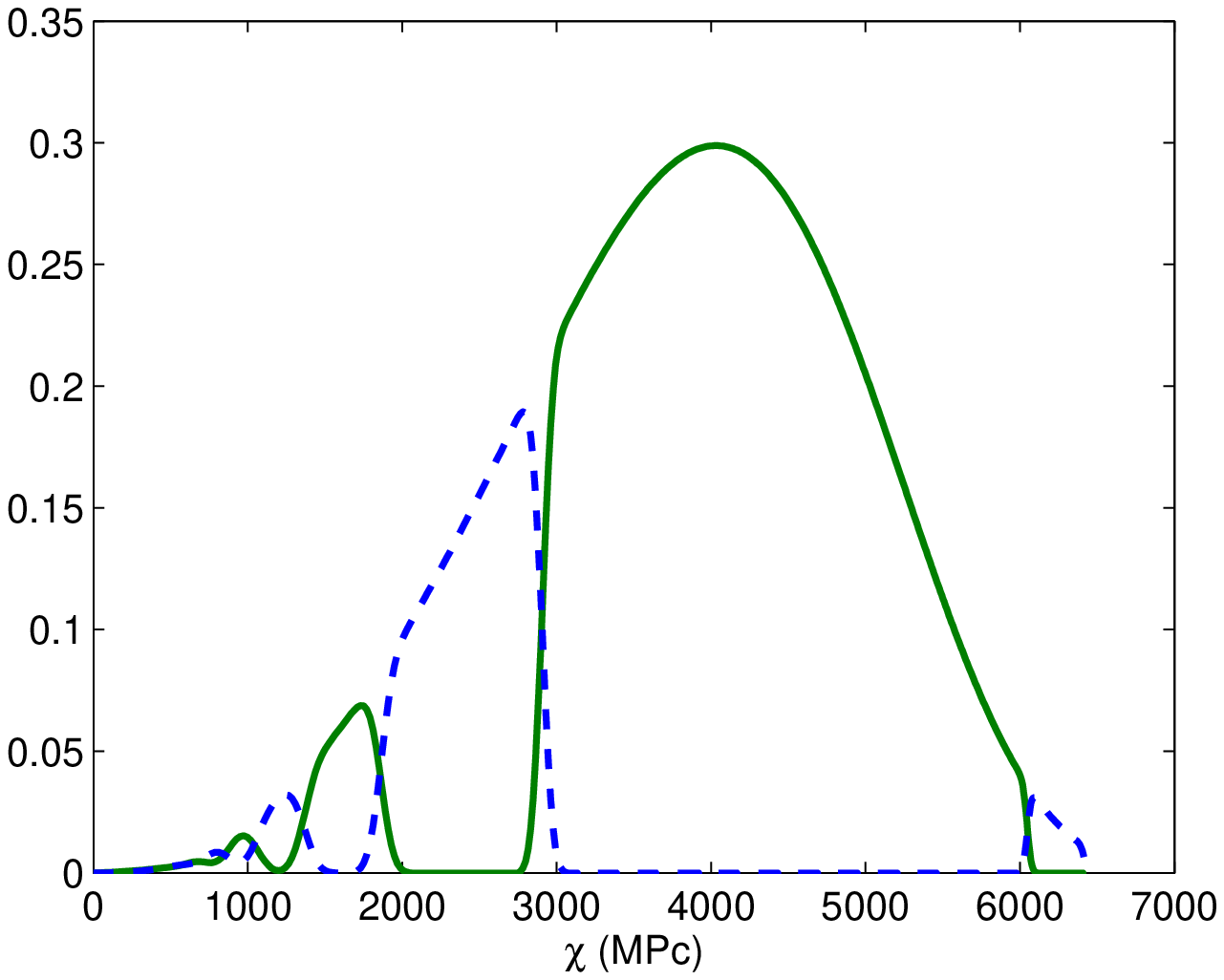}
\includegraphics[width=3.4in]{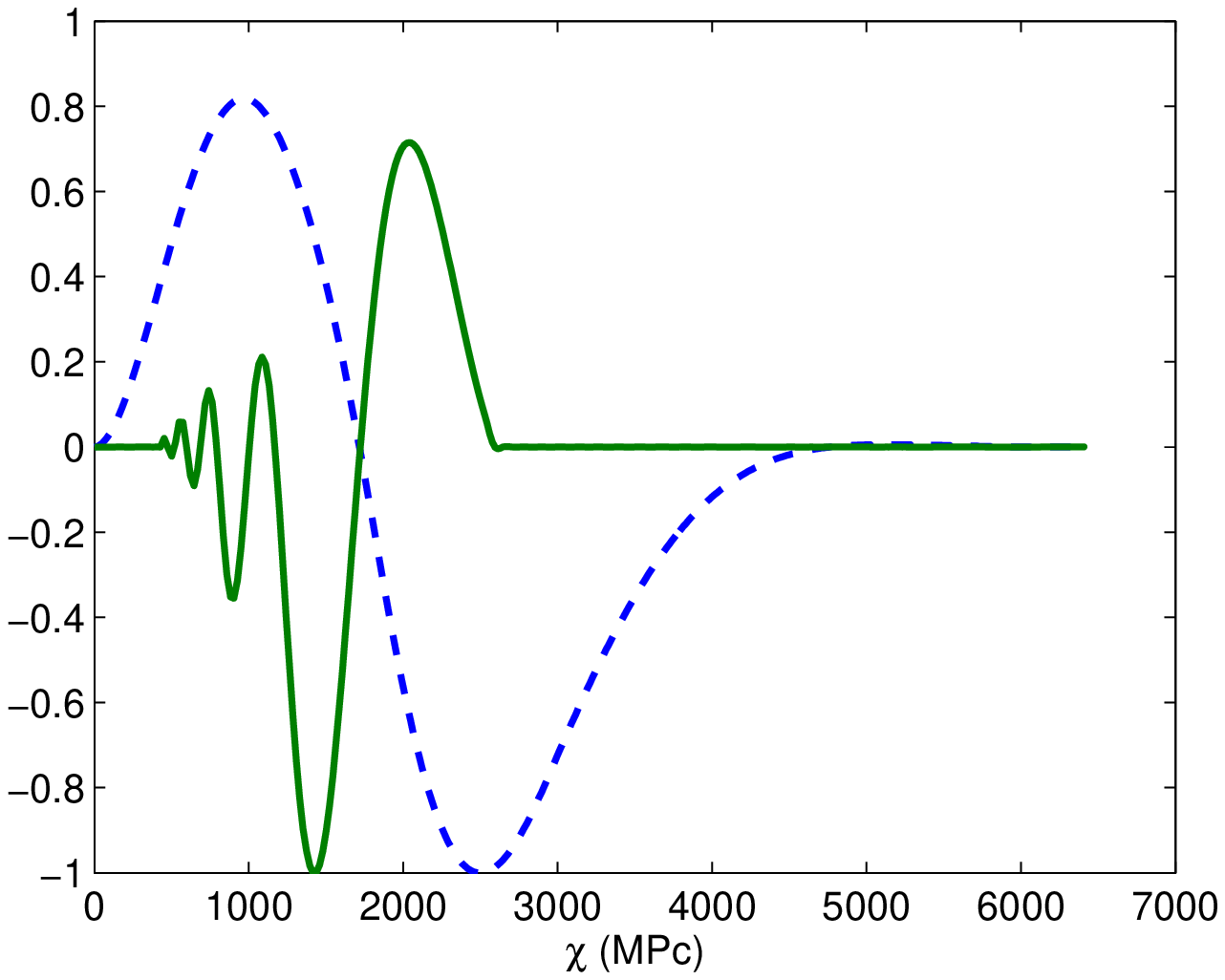}
\caption{\label{fig:100} $\ell=100$ (Top) The solid and dashed
lines represents the galaxy distribution for bins A and B
respectively. The uneven allocation of galaxies is undesirable as
it leads to a large sampling variance for bin B. (Bottom) The
solid line is the difference between $P(\frac{\ell}{r},\chi)$ and
a smoothed version. The dashed line is the effective lensing
efficiency, as defined in the text. In both diagrams, the vertical
axes have arbitrary normalisation.}
\end{figure}

It is important to note that due to the $\ell$-dependence of
$\epsilon$, the binning needs to be performed separately for each
group of multipoles evaluated.  Particular examples are
illustrated in the top panels of Figs.~\ref{fig:100} and
\ref{fig:200}.  Here we have assumed a photometric redshift
accuracy of $\Delta z \simeq 0.02$, simulated by convolving our
distribution with a Gaussian. This can be seen as a smoothing of
the bin edges.

\section{Bin Comparison}

We aim to devise a statistic which is sensitive to the
oscillation, whilst remaining invariant to the overall height of
the power spectrum. In doing so, this reduces the large cosmic
variance error generated by the smooth component of the power
spectrum. One problem we are faced with is that $g$, as defined in
(\ref{eq:lenseff}) and (\ref{eq:normal}), is a smooth function and
so oscillations in $P(\frac{\ell}{r},\chi)$ are undetectable. We
must therefore make use of the redshift bins as derived in the
previous section.

Hu \cite{1999ApJ...522L..21H} found that the straightforward
binning into high and low redshift galaxies still resulted in a
high level of correlation. Here we use this to our advantage. The
two intertwined bins share a great deal of structure and thus when
subtracting their signals, the cosmic variance contribution to the
error is greatly reduced.

\begin{figure} [t]
\includegraphics[width=3.4in]{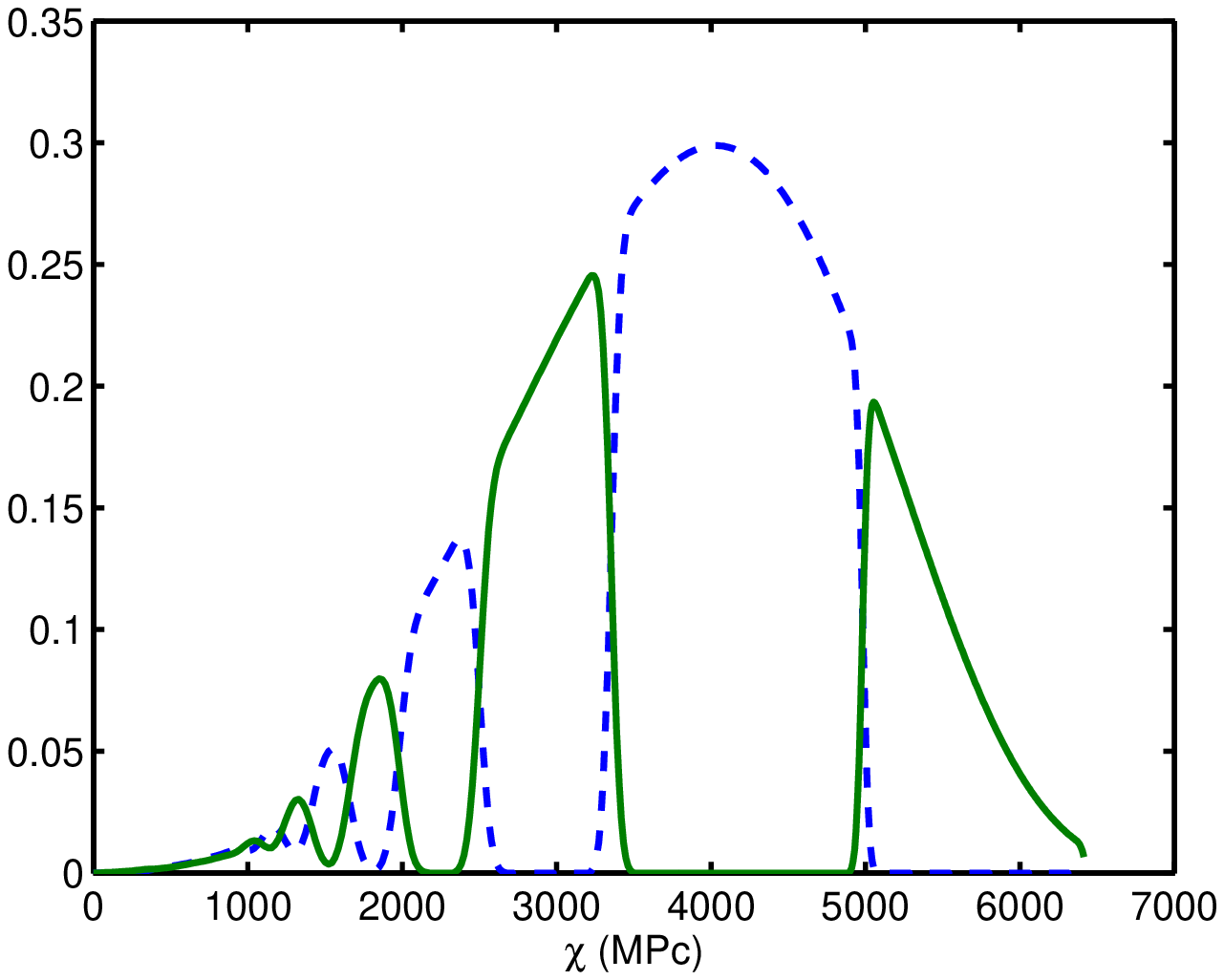}
\includegraphics[width=3.4in]{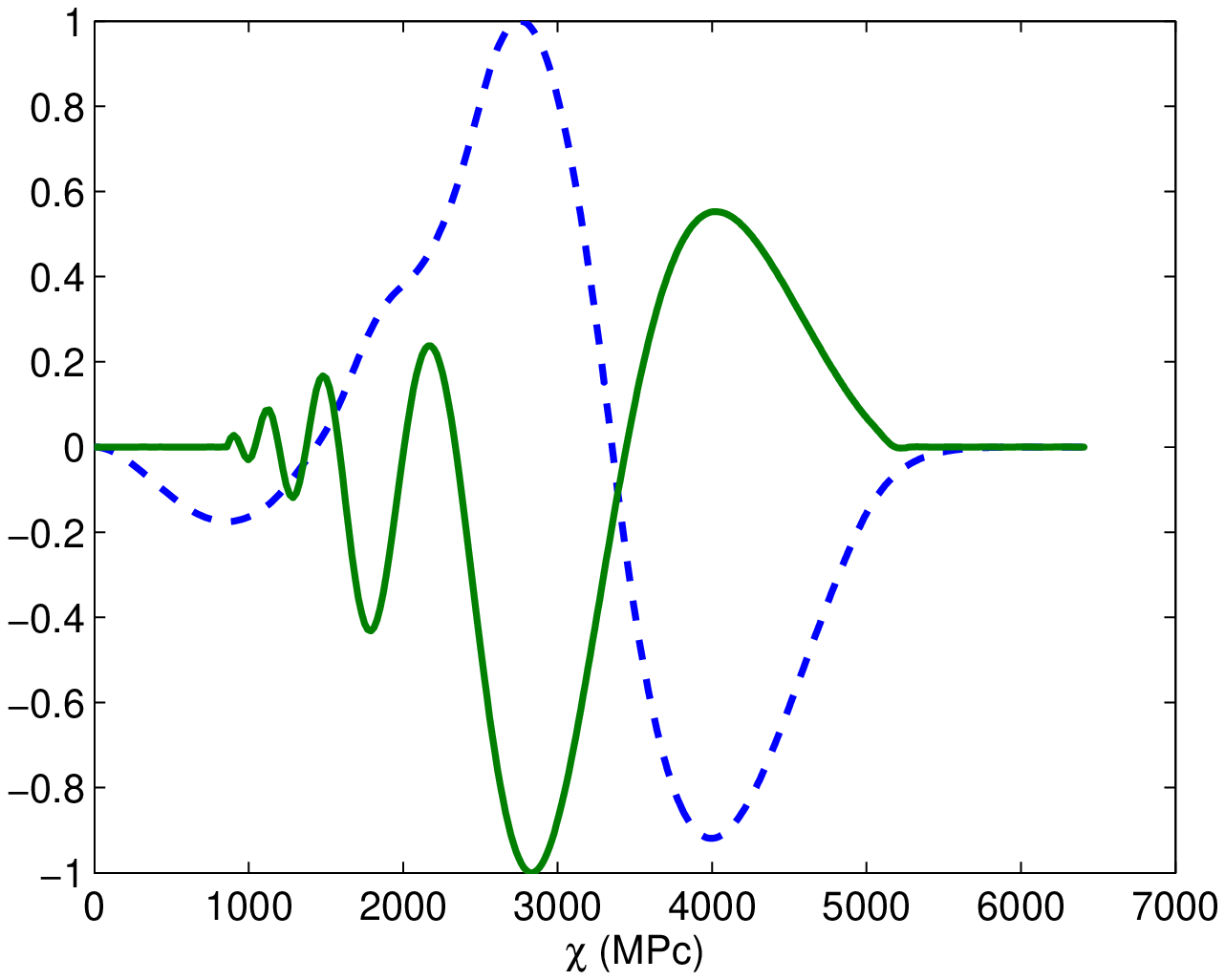}
\caption{\label{fig:200} The plots here are the same as in
Fig.~\ref{fig:100}, but now $\ell=200$.}
\end{figure}

To begin, we define the statistic $F(\ell)$ in terms of the
autocorrelations of the two bins from (\ref{eq:na}) and
(\ref{eq:nb}).

\begin{equation}
\label{eq:F} F(\ell) \equiv \Gamma C^{AA}_\ell - C^{BB}_\ell
\end{equation}

\noindent where the scaling parameter $\Gamma$ is given in terms of
the fiducial cosmological model.

\begin{equation}
\Gamma = \tilde{C}_\ell^{BB}/\tilde{C}_\ell^{AA}.
\end{equation}

From (\ref{eq:shear}) and (\ref{eq:F})  we have

\begin{align} \label{eq:Fint}
F(\ell)= & \frac{9}{16}\Big(\frac{H_0}{c}\Big)^4 \Omega_m^2 \times\\
& \nonumber \int^\chi_0 \Big[\frac{\displaystyle \Gamma
g^2_{A}(\chi)-g^2_{B}(\chi)}{\displaystyle
a^2r^2(\chi)}\Big]P(\frac{\ell}{r},\chi) \ud \chi .
\end{align}

The important component of this equation is the \emph{effective
  lensing efficiency}, given by $[\Gamma g^2_{A}(\chi)-g^2_{B}]$.
In stark contrast to the standard lensing efficiency from
\ref{fig:smooth}, this term can be seen to trace the baryonic
oscillations in Figs.~\ref{fig:100} and \ref{fig:200}. The top
plots demonstrate the $\ell$-dependent binning, allowing the
effective lensing efficiency to adapt to the changing position of
the wiggles, as depicted in the bottom graphs.

\section{Analysis}

There are a number of factors to consider when trying to maximise
the signal-to-noise ratio generated by a change in the amplitude of
oscillations. For example, uneven separation of galaxies into the
two bins leads to an enhanced sampling variance. We also desire the
production of an effective lensing efficiency which closely follows
the oscillations. Thus we adopt a brute-force approach to evaluate
the optimal phase $\phi$ for the binning function $\epsilon(\chi,
\ell)$ discussed in the previous section. This was found to be well
approximated by $\phi \simeq 0.26-0.0125\ell$.

Let us return to equation (\ref{eq:error}).  Within the brackets, the two
terms are attributed to cosmic variance and the sampling variance. The
former tends to dominate at lower multipoles since the shear signal is
greater.

For the full error expression of $F(\ell)$ we find

\begin{eqnarray}
\sigma_F=\sqrt{\sigma_A^2+\sigma_B^2-2r^2\sigma_A\sigma_B} \\
 \sigma_A = \Gamma
\sqrt{\frac{\displaystyle 2}{\displaystyle(2\ell +1)f_{sky}}}
(C^{AA}_\ell+\frac{\displaystyle\sigma^2_\gamma}{\displaystyle
2n_A}) \\
\sigma_B = \sqrt{\frac{\displaystyle 2}{\displaystyle(2\ell
+1)f_{sky}}}
(C^{BB}_\ell+\frac{\displaystyle\sigma^2_\gamma}{\displaystyle
2n_B}) \\
\sigma_{AB} = \sqrt{\frac{\displaystyle
2\Gamma}{\displaystyle(2\ell +1)f_{sky}}} (C^{AB}_\ell).
\end{eqnarray}

The correlation coefficient is given by
\begin{equation}
r^2 = \frac{\sigma_{AB}^2}{\sigma_A\sigma_B} ,
\end{equation}
\noindent where for brevity we have defined the terms $A \equiv \Gamma
C^{AA}_\ell$ and $B \equiv C^{BB}_\ell$. The covariance between
the terms A and B is denoted by $\sigma_{AB}^2$. We typically find
values of $r \sim 0.95$.

As an illustration, we use a power spectrum which has been
smoothed by interpolating between the oscillations.

In order to quantify the constraint, we parameterise the amplitude
of oscillations as $O$, such that $O=1$ in the standard case, and
$O=0$ for the smoothed model. When marginalising over various
cosmological parameters ($\Omega_m$,$n$,$w$,$h$,$\sigma_8$), we
must be careful to include the effect of altering $n(\chi)$ since
the bin selection must be made in redshift space. Adopting a
standard Fisher matrix approach, we find that the error does not
significantly degrade from marginalisation, due to the unique
functional form the signal imprints on $F(\ell)$. We have also
investigated the variation of $\sigma_F$ with cosmology, and
tested higher redshift errors ($\Delta z \simeq 0.1$), and found
these effects to be negligible. If the rms shear is increased by
$50\%$, constraints degrade by $\sim 30\%$.

For our fiducial survey, without using tomography, the constraints
on $O$ are very poor, as expected from Fig. \ref{fig:smooth}. We
find $\sigma_O=6$. However, by applying the approach outlined
above, the smooth power spectrum can be ruled out at the $2\sigma$
level. We therefore anticipate that surveys at the level of the
SKA \cite{2004NewAR..48.1063B}, are required before significant
results can be drawn.

\section{Discussion}

We have demonstrated how angular-dependent tomography can allow
forthcoming cosmic shear surveys to detect oscillations in the
matter power spectrum. Whilst this will never compete with
high-redshift galaxy surveys, it does provide a unique consistency
check. Cosmic shear and galaxy surveys probe the dark matter power
spectrum via completely independent routes, and so such a
complementary probe is welcome. It may also be of some assistance
in controlling systematics.

It is likely that the errors derived here may be an
\emph{overestimate}, due to improvements which could be made with
a more elaborate binning algorithm, and the lack of priors on
cosmological parameters. Additionally, by weighting the galaxies
(King \& Schneider \cite{2002A&A...396..411K}) we could improve
the extent to which $g$ tracks the oscillations in the power
spectrum, thereby enhancing the signal. However we anticipate that
this gain may be tempered by the additional shot noise introduced,
so its application could be limited to the lower multipoles where
cosmic variance remains dominant.

The erosion of the baryonic features at small scales, by nonlinear
structure, is not expected to affect our results since the signal
is dominated by the strength of the first two oscillations.

``Precision cosmology" and a ``concordance model" are
oft-mentioned phrases, but perhaps a little premature
\cite{2003Sci...299.1532B}. The advent of cosmic shear surveys
which cover a significant fraction of the sky is likely to deliver
both.

\begin{acknowledgments}
The author would like to thank S.Bridle for helpful comments, and is
grateful for the support from Trinity College.
\end{acknowledgments}
\bibliography{bibs}
\end{document}